\documentclass[pre]{revtex4}

\usepackage{amsmath}
\usepackage{amsfonts}
\usepackage{hyperref}
\usepackage{graphicx}

\usepackage[T1]{fontenc}

\newcommand{\dd}{\mathrm{d}}%
\newcommand{\ii}{\mathrm{i}}
\newcommand{\bfrho}{\boldsymbol{\rho}}

\newcommand{\bfr}{\boldsymbol{r}}
\newcommand{\bfR}{\boldsymbol{R}}

\newcommand{\bfxi}{\boldsymbol{\xi}}

\newcommand{\bfk}{\boldsymbol{k}}

\newcommand{\bfp}{\boldsymbol{p}}

\DeclareMathOperator{\spn}{span}

\newcommand{\bra}{\langle}
\newcommand{\ket}{\rangle}

\begin{document}

\title{On the Twistability of Partially Coherent, Schell-model Sources}

\author{Riccardo Borghi}
\affiliation{Departimento di Ingegneria Civile, Informatica e delle Tecnologie Aeronautiche, 
Universit\`{a} ``Roma Tre'', Via Vito Volterra 62, I-00146 Rome, Italy}

\begin{abstract}
In this paper, the problem of assessing the twistability of a given bona fide cross-spectral density is tackled for 
the class of Schell-model sources, whose shift-invariant degree of coherence is represented by a real 
and symmetric function,
{denoted as} 
 $\mu(-\bfr)=\mu(\bfr)$. By employing an abstract operatorial language, the 
problem of determining the highly degenerate spectrum of a twisted operator $\hat W_u$ is 
addressed through a modal analysis based on {the}
 complete knowledge of the spectrum of the 
{\em sole} twist operator $\hat T_u$, as found 
by
R. Simon and N. Mukunda.  [J. Opt. Soc. Am. A \textbf{15,} 1361 (1998)].
To this 
end, the evaluation of the complete tensor of the matrix elements $\bra n',\ell'|\hat W_u|n,\ell\ket$  is carried out within the framework
of the so-called
{\em extended Wigner distribution function}
, a concept recently introduced by
M. {VanValkenburgh} [J. Mod. Opt. \textbf{55,}  3537 - 3549 (2008)]
. As a nontrivial application of the algorithm developed here, the analytical determination of the spectrum of saturated twisted 
astigmatic Gaussian Schell-model sources is also presented.
\end{abstract}

\maketitle

\section{Introduction}
\label{Sec:Intro}

{One of the most charming} and still not completely explored topics of classical coherence theory is, without a~doubt, that which deals with so-called 
{\em {twisted sources} 
} 
and the beams they generate. 
In a celebrated 1993 paper~\cite{Simon/Mukunda/1993}, Simon and Mukunda opened up ``a new dimension'' in coherence theory~\cite{Fiberg/Tervonen/Turunen/1994} through the introduction of 
{a} 
``genuinely two-dimensional''~\cite{Simon/Mukunda/1993} axially symmetric twist phase to be imposed on ``ordinary'' Gaussian Schell-model (GSM) beams. In~this way, the~class of the celebrated  {\em Twisted Gaussian Schell-model} (TGSM henceforth) beams was born.
The multiplication of the cross-spectral density (CSD henceforth) of a standard GSM source by the new chiral phase term
produced a CSD whose striking properties continue to be explored, both theoretically and experimentally, following~the original Simon and Mukunda paper~\cite{Simon/Mukunda/1993}, as well as a series of important follow-up works~\cite{Simon/Sundar/Mukunda/1993,Simon/Sundar/Mukunda/1993b,Fiberg/Tervonen/Turunen/1994}. TGSM beams have received considerable attention in the last thirty years, as evidenced by the countless list of papers they have inspired. 
Simon and Mukunda also drew attention in Ref.~\cite{Simon/Mukunda/1998} to a fundamental theoretical topic:  ``sticking'' a twist phase to a typical {bonafide} 
 untwisted CSD does not, in~general, result in a {bonafide} 
 twisted CSD. 
Even when a simple GSM source is to be twisted, the~strength of the twist phase must satisfy severe numerical limitations, which are imposed by the coherence features of the {\em sole} spectral degree of the untwisted GSM source~\cite{Simon/Mukunda/1993}. Important results were found 
in~\cite{Simon/Mukunda/1998b} for the class of all shape-invariant anisotropic GSM
(AGSM henceforth) beams. 
{In a sense, that paper gave a definite answer to the key question of how and when a twist phase acts as an operator that can map untwisted {bonafide} 
 CSDs into twisted {bonafide} 
 CSDs in the broader context of Gaussian optics}~\cite{Simon/Mukunda/1998,Simon/Mukunda/2000}.

In 2015, what it could be called  the ``twist mapping problem'' was addressed for a considerably larger class, with~respect to AGSM sources, of~{bonafide}
, Schell-model CSDs~\cite{Borghi/Gori/Santarsiero/Guattari/2015}. 
In order to introduce our readers the main definitions and notations of the present paper, 
let us consider the mapping problem from as general a perspective as possible.
Given a {bonafide} 
 CSD associated with a planar partially coherent source, say $W(\bfr_1,\bfr_2)$,  we shall ask under what conditions the new CSD, 
say $W_u(\bfr_1,\bfr_2)$, defined by
\begin{equation}
\label{Eq:TwistedGeneraleCSD.01}
\begin{array}{l}
\displaystyle
W_u(\bfr_1,\bfr_2)\,=\,W(\bfr_1,\bfr_2)\,\exp(-\ii u \bfr_1\times\bfr_2)\,,\qquad \qquad u>0\,,
\end{array}
\end{equation}
still represents a {bonafide} 
 CSD. Here, $u$ is the twist strength (which will be taken as positive), 
while the symbol $\times$ represents the $z$-component of the cross product between the transverse 
position vectors $\bfr_1$ and $\bfr_2$ across the source plane, which is orthogonal to the $z$-axis.
In other words, the~symbol $\bfr_1\times\bfr_2$ is a ``notational shortcut'' for the mixed product $\bfr_1\times\bfr_2\cdot\hat\bfk$,
with $\hat\bfk$ being the unit vector orthogonal to the source plane. 
In Ref.~\cite{Borghi/Gori/Santarsiero/Guattari/2015}, the~mapping problem was formally solved for the whole class of CSDs
of the following, Schell-model type: 
\begin{equation}
\label{Eq:TwistedGeneraleCSD}
\begin{array}{l}
\displaystyle
W(\bfr_1,\bfr_2)\,=\,\tau^*(\bfr_1)\,\tau(\bfr_2)\,\mu(|\bfr_1-\bfr_2|)\,,
\end{array}
\end{equation}
where $\tau(\cdot)$ denotes the transmission function of an arbitrary (complex) amplitude filter 
and $\mu(\cdot)$ represents a shift-invariant, spectral degree of coherence endowed with 
{\em radial symmetry}. 
The necessary and sufficient condition for $W$ to be {bonafide} 
 follows from Bochner's theorem, 
which implies that the Fourier transform of $\mu$ must 
 be identically non-negative.
The key idea of Ref.~\cite{Borghi/Gori/Santarsiero/Guattari/2015} to solve the twist mapping problem was based on the fact that 
the twisted version of the CSD in Equation~(\ref{Eq:TwistedGeneraleCSD}) will be {bonafide} 
 if and only if the 
corresponding {\em uniform} CSD, obtained by removing the complex filter $\tau(\bfr)$, i.e.,
\begin{equation}
\label{Eq:UniformTwistedGeneraleCSD}
\begin{array}{l}
\displaystyle
W_u(\bfr_1,\bfr_2)\,=\,
\mu(|\bfr_1-\bfr_2|)\,\exp\left(-\mathrm{i}u\,\bfr_1\times\bfr_2\right)
\,,\qquad\qquad u >0\,,
\end{array}
\end{equation}
is {bonafide} 
 too. 
The solution of the twist problem in Ref.~\cite{Borghi/Gori/Santarsiero/Guattari/2015} also followed from the conjecture that the Wolf coherent modes~\cite{Mandel/Wolf} of the uniform kernel $W_u(\bfr_1,\bfr_2)$ in 
Equation~(\ref{Eq:UniformTwistedGeneraleCSD}) coincide with the so-called Laguerre–Gauss (LG henceforth) beams, say $\Phi_{j,m}(\bfr)$, defined as~\cite{Mandel/Wolf}
\begin{equation}
\label{Eq:ModalExpansionUniformTGSM.3}
\begin{array}{l}
\displaystyle
\Phi_{j,m}(\bfr)\,=\,\sqrt{\frac u\pi}\,\left[\frac{(j-|m|)!}{(j+|m|)!}\right]^{\frac 12}\,
(r\sqrt {u})^{2|m|}\,\exp(\mathrm{i}2m\varphi)\,L^{2|m|}_{j-|m|}(ur^2)\,\exp\left(-\frac{ur^2}2\right)\,.
\end{array}
\end{equation}
Here
, $j=0,1/2,1,\ldots$ represents the infinite sequence of semi-integer and integer non-negative numbers, while 
$m=-j,-j+1,\ldots,j-1,j$. In~this way, it is sure that both radial  ($j-|m|$) and angular ($2|m|$) indices of 
Laguerre polynomials take on non-negative {\em integer} values. 
Once the modal structure was conjectured, the~complete spectrum of $W_u$
was found. Accordingly, in~order to assess the non-negative definiteness of the twisted source,
it was sufficient (but also necessary) to impose each eigenvalue to be non-negative~\cite{Borghi/Gori/Santarsiero/Guattari/2015}. 
That is~all. 

The results presented in Ref.~\cite{Borghi/Gori/Santarsiero/Guattari/2015} are based on a conjecture that was not 
fully justified from a mathematical viewpoint.
One of the aims of the present paper is to provide such a rigorous justification.
In doing so, we shall address the twist mapping problem for a more general class of Schell-model
CSDs than those defined in Equation~(\ref{Eq:TwistedGeneraleCSD}), namely those for 
which the spectral degree of coherence $\mu$ is not necessarily a radial function, but rather~satisfies 
\begin{equation}
\label{Eq:MiuSymmetry}
\begin{array}{l}
\displaystyle
\mu(-\bfr)\,=\,\mu(\bfr)\,,\qquad\qquad\qquad \bfr \in \mathbb{R}^2\,,
\end{array}
\end{equation}
which, due to the mandatory Hermiticity of $W_u$, implies $\mu \in \mathbb{R}$.

Our strategy is based on the approach outlined in~\cite{Simon/Mukunda/1998}:  a typical CSD $W(\bfr_1,\bfr_2)$ 
will be thought of as the position representation of a Hermitian operator, say  $\hat W$, 
acting on the Hilbert space $\mathcal{H}$ of squared integrable wavefunctions. 
To help our readers, it is worth giving a brief introduction of  the formalism that will be used throughout the present paper.
According to the notations of Ref.~\cite{Simon/Mukunda/1998}, let $|\psi\ket$ be the abstract state (ket) which corresponds, in~the position representation, to~the wavefunction $\psi(\bfr)$, i.e.,
\begin{equation}
\label{Eq:Notations.1}
\begin{array}{l}
\displaystyle
\psi(\bfr)\,=\,\bra \bfr | \psi \ket\,,
\end{array}
\end{equation}
where, as~usual, the~symbol $|\bfr\ket$ denotes the eigenket of the position operator $\hat\bfr$, defined as
\begin{equation}
\label{Eq:Notations.2}
\begin{array}{l}
\displaystyle
\hat{\bfr} |\bfr\ket \,=\,\bfr | \bfr \ket\,.
\end{array}
\end{equation}
In this way, the~connection between the CSD $W(\bfr_1,\bfr_2)$ and the associated 
 abstract operator $\hat W$ will be given
through its matrix elements, according to
\begin{equation}
\label{Eq:Notations.3}
\begin{array}{l}
\displaystyle
W(\bfr_1,\bfr_2)\,=\,\bra{\bfr_1} |\hat W|\bfr_2\ket\,.
\end{array}
\end{equation}
The operator $\hat W$ acts on a ket $|\psi\ket \in \mathcal{H}$ to generate another element of $\mathcal{H}$, which is denoted by
$\hat W |\psi\ket$, whose position representation turns out to be
\begin{equation}
\label{Eq:Notations.4}
\begin{array}{l}
\displaystyle
\bra \bfr |\hat W|\psi \ket\,=
\int_{\mathbb{R}^2}\,\dd^2\rho\,
\bra \bfr |\hat W| \bfrho \ket \bra \bfrho |\psi \ket\,=\,
\int_{\mathbb{R}^2}\,\dd^2\rho\,
W(\bfr,\bfrho) \psi(\bfrho)\,,
\end{array}
\end{equation}
where the formal completeness condition, i.e.,
\begin{equation}
\label{Eq:Notations.5}
\begin{array}{l}
\displaystyle
\mathbb{I}\,=\,
\int_{\mathbb{R}^2}\,\dd^2\rho\,
| \bfrho \ket \bra \bfrho |\,,
\end{array}
\end{equation}
with $\mathbb{I}$ being the identity operator, has been used.
The other ingredient of our mathematical recipes is the commutation relationship between the 
two operators.
In particular, in~Ref.~\cite{Borghi/Gori/Santarsiero/Guattari/2015} it was shown that, for~radial spectral degrees of coherence,
the operator $\hat W_u$ associated with the twisted CSD defined in Equation~(\ref{Eq:UniformTwistedGeneraleCSD}) and the operator, say $\hat T_u$, associated with the sole twist kernel $T_u(\bfr_1,\bfr_2)=\exp(-\ii u \bfr_1\times\bfr_2)$, {\em do commute}, i.e.,
\begin{equation}
\label{Eq:Notations.5.1}
\begin{array}{l}
\displaystyle
\hat W_u\,\hat T_u\,=\,\hat T_u\,\hat W_u\,.
\end{array}
\end{equation}
It is not difficult to prove that Equation~(\ref{Eq:Notations.5.1}) holds 
for any  non-radial function $\mu(\bfr)$ that satisfies Equation~(\ref{Eq:MiuSymmetry}).
For the sake of simplicity, the~proof has been confined to Appendix~\ref{App:Commutation}.


\section{Commuting Operators and Spectral~Degeneration}
\label{Sec:COSD}

An important achievement of Ref.~\cite{Simon/Mukunda/1998} 
has to do with the spectral theorem for the twist operator $\hat T_u$.
In particular, Simon and Mukunda proved that
\begin{equation}
\label{Eq:COSD.1}
\begin{array}{l}
\displaystyle
\hat T_u |{j,m}\ket \,=\,\dfrac{2\pi}u\,(-1)^{j-m}\,|{j,m}\ket\,,\qquad\qquad \left({{j=0,1/2,1,\ldots}\atop{m=-j,-j+1,\ldots,j-1,j}}\right),
\end{array}
\end{equation}
where the position representation of the eigenket $|j,m\rangle$  coincides with the LG beam of Equation~(\ref{Eq:ModalExpansionUniformTGSM.3}), i.e.,
\begin{equation}
\label{Eq:COSD.1.1}
\begin{array}{l}
\displaystyle
\bra \bfr |{j,m}\ket \,=\,\Phi_{j,m}(\bfr)\,,\qquad\qquad \left({{j=0,1/2,1,\ldots}\atop{m=-j,-j+1,\ldots,j-1,j}}\right).
\end{array}
\end{equation}
From Equation~(\ref{Eq:COSD.1}), it turns out that the spectrum of the twist operator $\hat T_u$ is highly degenerate. 
It is worth introducing two auxiliary integer parameters, say $n$ and $\ell$, defined as $n=j+m$, $\ell=j-m$, so that 
\begin{equation}
\label{Eq:COSD.2}
\begin{array}{l}
\displaystyle
|{j,m}\ket \,=\,\left|\dfrac{n+\ell}2,\dfrac{n-\ell}2\right\rangle\,,\qquad\qquad n,\ell=0,1,2\ldots
\end{array}
\end{equation}
In the following, given the one-to-one relation $(j,m)\Longleftrightarrow (n,\ell)$, the~eigenket $|{j,m}\ket$ will be denoted  by 
$|{n,\ell}\ket$ in place of Equation~(\ref{Eq:COSD.2}), without~lack of clarity.
With these notations, Equation~(\ref{Eq:COSD.1}) will be recast as 
\begin{equation}
\label{Eq:COSD.3}
\begin{array}{l}
\displaystyle
\hat T_u |n,\ell\ket \,=\,\dfrac{2\pi}u\,(-1)^{\ell}\,|n,\ell\ket\,,\qquad\qquad n,\ell=0,1,2,\ldots
\end{array}
\end{equation}
The high spectral degeneration of the twist operator represents one of the principal technical problems to be addressed.
In particular, the~spectrum of the twist operator consists in only two eigenvalues, namely $2\pi/u$ and $-2\pi/u$,
each of them infinitely degenerate. Accordingly, the~Hilbert space $\mathcal{H}$
can be thought of as the union of two subspaces, say $\mathbb{V}_\pm$, defined as follows:
\begin{equation}
\label{Eq:COSD.3.1}
\left\{
\begin{array}{l}
\displaystyle
\mathbb{V}_{+}\,=\,\spn\{|n,2k\ket\}_{n,k=0}^\infty\,,\\
\\
\mathbb{V}_{-}\,=\,\spn\{|n,2k+1\ket\}_{n,k=0}^\infty\,,\\
\end{array}
\right.
\end{equation}
where each of them is generated by all eigenkets corresponding to the {\em same} value of the eigenvalue. 
It is a well-known fact that, since the operators $\hat W_u$ and $\hat T_u$ commute, the~ket 
$\hat W_u |n,\ell\ket$ must itself be an eigenket of $\hat T_u$
corresponding to the eigenvalue $\dfrac{2\pi}u (-1)^\ell$. In~fact, 
\begin{equation}
\label{Eq:COSD.4}
\begin{array}{l}
\displaystyle
\hat T_u\hat W_u |n,\ell\ket \,=\,
\hat W_u\hat T_u |n,\ell\ket \,=\,
\dfrac{2\pi}u(-1)^\ell\,\hat W_u |n,\ell\ket \,,\qquad\qquad n,\ell=0,1,2,\ldots
\end{array}
\end{equation}
In Ref.~\cite{Borghi/Gori/Santarsiero/Guattari/2015}, it was conjectured that, for~any radial degree of coherence $\mu$, the~
modes of the twisted CSD $W_u$ of Equation~(\ref{Eq:UniformTwistedGeneraleCSD}) coincide with the modes of $T_u$.
One of the scopes of the present paper is also to provide a mathematical justification for it, {which will be presented in 
Section~\ref{Sec:ProvingRadialMu}}.
In order to better clarify the terms of the problem, consider the following case, namely $\mu(x,y)=\exp(-x^2-\chi y^2)$,
where $\chi$ represents a real parameter running within the interval $[1,\infty[$ (the choice $\chi=1$
corresponds to a Gaussian spectral degree of coherence, as~for example  in the classical GSM case). 
Consider then the fundamental state $|0,0\ket$, such~that 
\begin{equation}
\label{Eq:COSD.4.1}
\begin{array}{l}
\displaystyle
\bra \bfr |0,0\ket=\Phi_{0,0}(\bfr)\,=\,\sqrt{\dfrac u\pi}\,\exp\left(-u\dfrac{x^2+y^2}2\right)\,.
\end{array}
\end{equation}
Then, it turns out 
 ({we performed this evaluation with the help the latest release (14.1) of {\em Mathematica}})

\vspace{-12pt}
\centering

\begin{equation}
\label{Eq:COSD.4.0.0.1}
\begin{array}{l}
\displaystyle
\bra \bfr |
\hat W_u |0,0\ket\,=\,\\
\\
\,=\,
\dfrac{2\pi}{\sqrt{(u+2)(u+2\chi)}}\,\sqrt{\dfrac u\pi}\exp\left[-\dfrac{u(x-\ii y)(u^2(x+iy)^2+4(x+\ii y)+4u (x+\ii \chi y))}{2(2+u)(u+2\chi)}\right]\,,
\end{array}
\end{equation}
and it is not difficult to check that such a wavefunction corresponds to  an eigenket of the twist kernel $\hat T_u$, with an ~eigenvalue equal to 
$2\pi/u$. Moreover, in~the limit of $\chi\to 1^+$ (radial symmetry)
the state $\hat W_u |0,0\ket$ becomes proportional to $|0,0\ket$, as~conjectured in~\cite{Borghi/Gori/Santarsiero/Guattari/2015}. More~precisely, 
\begin{equation}
\label{Eq:COSD.4.0.0.1.2.0}
\begin{array}{l}
\displaystyle
\lim_{\chi\to 1}\,\hat W_u |0,0\ket\,=\,\dfrac{2\pi}{u+2}\,|0,0\ket\,.
\end{array}
\end{equation}
If $\chi>1$, the~state $\hat W_u |0,0\ket$ is expected to belong to the subspace $\mathbb{V}_+$ defined in \mbox{Equation~(\ref{Eq:COSD.3.1})}. 
Accordingly, it is natural to write
\begin{equation}
\label{Eq:COSD.4.0.0.1.2.0.1}
\begin{array}{l}
\displaystyle
\hat W_u |0,0\ket\,=\,\sum^\infty_{n'=0}\,\sum^\infty_{\ell'=0}\,
\bra n',\ell'|\hat W_u |0,0\ket\,|n',\ell'\ket\,,
\end{array}
\end{equation}
where, since $|0,0\ket \in \mathbb{V}_+$, only {\em even} values of the index $\ell'$ would be involved into
the double series in Equation~(\ref{Eq:COSD.4.0.0.1.2.0.1}). As~it will  be discussed in Section~\ref{Subsec:ATGS}, 
the possibility of finding such a representation would allow, in~principle, to~solve the degenerate eigenvalue problem
for the $\hat W_u$ operator and, consequently, to~assess its (semi)positive~definiteness.
 
The above example should be enough to convey to our readers at least the complexity 
behind the ``twistability problem,'' i.e.,~to find the necessary and sufficient conditions
for the kernel $W_u(\bfr_1,\bfr_2)$, defined as
\begin{equation}
\label{Eq:UniformTwistedGeneraleCSDCSD}
\begin{array}{l}
\displaystyle
W_u(\bfr_1,\bfr_2)\,=\,
\mu(\bfr_1-\bfr_2)\,\exp\left(-\mathrm{i}u\,\bfr_1\times\bfr_2\right)
\,,\qquad\qquad u >0\,,
\end{array}
\end{equation}
to represent a {bonafide} 
 CSD, under~the sole hypothesis~(\ref{Eq:MiuSymmetry}). 

The strategy pursued in the present work is to extract the complete representation of the twisted operator $\hat W_u$ in terms of
the orthonormal basis of the twist operator $\hat T_u$, in a similar manner to Equation~(\ref{Eq:COSD.4.0.0.1.2.0.1}),  but now for a typical
state $|n,\ell\ket$. To~this end, the~main technical problem to be solved is the evaluation of the typical 
matrix element  $\bra n',\ell'|\hat W_u|n,\ell\ket$. In~the next section, it will be shown that such a matrix element can be expressed in terms 
of the inner product between the degree of coherence $\mu(\bfr)$ and a suitable LG mode of Equation~(\ref{Eq:ModalExpansionUniformTGSM.3}).
To this end, the~results of an important paper, published in 2008 by 
{VanValkenburgh}~\cite{Valkenburgh/2008}
, will be employed.
In spite of the strong mathematical character of all the steps, we prefer not to give them the 
aspect of an appendix, but~rather, due to their key role, to~arrange them into a section 
that represents the technical core of the present~paper. 

\section{Evaluation of \boldmath{$\hat W_u$}'s Matrix~Elements}
\label{Sec:Integral}

The typical matrix element
$\langle n',\ell' | \hat W_u |n,\ell\rangle$ is first explicited as a 4-dimensional integral, i.e.,~
\begin{equation}
\label{Eq:LambdaInnerProduct}
\begin{array}{l} 
\displaystyle
\langle n',\ell' | \hat W_u |n,\ell\rangle\,=\,
\int_{\mathbb{R}^2}\,\int_{\mathbb{R}^2}\,
\dd^2 r_1\,\dd^2r_2\,
\langle n',\ell'|\bfr_1\ket\,\bra\bfr_1 | \hat W_u |\bfr_2\ket\,\bra\bfr_2 |n,\ell\rangle\,=\,\\
\\
\displaystyle
\,=\,\int_{\mathbb{R}^2}\,\int_{\mathbb{R}^2}\,
\dd^2 r_1\,\dd^2r_2\,
{W}_u(\bfr_1,\bfr_2)\,
\Phi^*_{j',m'}(\bfr_1)\,\Phi_{j,m}(\bfr_2)\,,
\end{array}
\end{equation}
where the pairs $(j,m)=\left(\frac{n+\ell}2,\frac{n-\ell}2\right)$ and $(j',m')=\left(\frac{n'+\ell'}2,\frac{n'-\ell'}2\right)$ have been re-introduced for the sake of convenience.
In order to evaluate the integral in Equation~(\ref{Eq:LambdaInnerProduct}), the~following properties of LG modes 
will also be employed:
\begin{equation}
\label{Eq:LambdaInnerProduct.0.1}
\left\{
\begin{array}{l} 
\displaystyle
\Phi^*_{j,m}(\bfr)\,=\,\Phi_{j,-m}(\bfr)\,,\\
\\
\Phi_{j,m}(\bfr)\,=\,(-1)^{2 m}\,\Phi_{j,m}(-\bfr)\,,
\end{array}
\right.
\end{equation}
where $2m \in \mathbb{Z}$ and, similarly for~the pair $(j',m')$. 
On substituting from the second row of Equation~(\ref{Eq:LambdaInnerProduct.0.1}) into 
Equation~(\ref{Eq:LambdaInnerProduct}) and on letting $\bfr=\bfr_1-\bfr_2$, simple algebra gives

\begin{equation}
\label{Eq:LambdaInnerProduct.1}
\begin{array}{l} 
\displaystyle
\langle n',\ell' | \hat W_u |n,\ell\rangle\,=\,(-1)^{2 m}\,
\int\,
\dd^2 r\,\mu(\bfr)\,
\int\,\dd^2r_2\,
\exp(-\mathrm{i} u \bfr\times\bfr_2)\,
\Phi^*_{j',m'}(\bfr+\bfr_2)\,
\Phi_{j,m}(-\bfr_2)\,,
\end{array}
\end{equation}
where the symbol $\displaystyle\int$ stands for $\displaystyle\int_{\mathbb{R}^2}$ henceforth.
The inner integral in Equation~(\ref{Eq:LambdaInnerProduct.1}) can be recast as a so-called {\em extended Wigner distribution function} (EWDF henceforth), a~concept proposed 
in {VanValkenburgh}'s paper~\cite{Valkenburgh/2008}. 
To this end, a~new integration variable, say $\bfrho=\bfr+2\bfr_2$, is introduced in place of $\bfr_2$, so that we have 
$-\bfr_2=\dfrac\bfr 2-\dfrac\bfrho 2$ and $\bfr+\bfr_2=\dfrac\bfr 2+\dfrac\bfrho 2$
which, after~being substituted into Equation~(\ref{Eq:LambdaInnerProduct.1}), give both 
\begin{equation}
\label{Eq:LambdaInnerProduct.3}
\begin{array}{l} 
\displaystyle
\int\,\dd^2r_2\,
\exp(-\mathrm{i} u \bfr\times\bfr_2)\,
\Phi^*_{j',m'}(\bfr+\bfr_2)\,\Phi_{j,m}(-\bfr_2)\,=\,\\
\\
\displaystyle
\,=\,\dfrac{1}4\,\int\,\dd^2\rho\,\exp\left(-\mathrm{i} \frac u2 \bfr\times\bfrho\right)\,
\Phi^*_{j',m'}\left(\dfrac\bfr 2\,+\,\dfrac\bfrho 2\right)\,
\Phi_{j,m}\left(\dfrac\bfr 2\,-\,\dfrac\bfrho 2\right)\,.
\end{array}
\end{equation}
It must be understood 
 how the $\bfrho$-integral turns out to be {\em independent of} the twist parameter $u$. 
To this end, it is sufficient to change the integration variable $\bfrho$ into the new variable
$\bfxi=\sqrt u\,\bfrho$, again taking Equation~(\ref{Eq:ModalExpansionUniformTGSM.3}) into account. 
Accordingly, from~Equation~(\ref{Eq:LambdaInnerProduct.3}) we have
\begin{equation}
\label{Eq:LambdaInnerProduct.4.1}
\begin{array}{l} 
\displaystyle
\int\,\dd^2r_2\,
\exp(-\mathrm{i} u \bfr\times\bfr_2)\,
\Phi^*_{j',m'}(\bfr+\bfr_2)\,\Phi_{j,m}(\bfr_2)\,=\,\\
\\
\displaystyle
\,=\,
\dfrac{(-1)^{2m}}4\,
\int\,\dd^2\bfxi\,
\exp\left(-\mathrm{i} \bfR \times\bfxi\right)\,
\bar{\Phi}^*_{j',m'}\left(\bfR\,+\,\dfrac\bfxi 2\right)\,
\bar{\Phi}_{j,m}\left(\bfR\,-\,\dfrac\bfxi 2\right)\,,
\end{array}
\end{equation}
where the dimensionless vector $\bfR=\dfrac{\bfr\sqrt u}2$ has been introduced and 
the symbol $\bar{\Phi}_{j,m}(\bfr)$ stands for $\left.\Phi_{j,m}(\bfr)\right|_{u=1}$. 
The subsequent step consists of moving the complex conjugation from the pair $(j',m')$
to the pair $(j,m)$ by using the first part 
of Equation~(\ref{Eq:LambdaInnerProduct.0.1}), which gives
\begin{equation}
\label{Eq:LambdaInnerProduct.4.2}
\begin{array}{l} 
\displaystyle
\int\,\dd^2r_2\,
\exp(-\mathrm{i} u \bfr\times\bfr_2)\,
\Phi^*_{j',m'}(\bfr+\bfr_2)\,\Phi_{j,m}(\bfr_2)\,=\,\\
\\
\displaystyle
\,=\,
\dfrac{(-1)^{2m}}4\,
\int\,\dd^2\bfxi\,
\exp\left(-\mathrm{i} \bfR \times\bfxi\right)\,
\bar\Phi^*_{j,-m}\left(\bfR\,-\,\dfrac\bfxi 2\right)\,,
\bar\Phi_{j',-m'}\left(\bfR\,+\,\dfrac\bfxi 2\right)\,.
\end{array}
\end{equation}
The mathematical structure of the last $\bfxi$-integral is the 
EWDF of the pair $\{\bar\Phi_{j,-m}\,,\bar\Phi_{j',-m'}\}$. In~fact, 
the quantity $\bfR\times\bfxi$ can be explicited as a mixed product as follows:
\begin{equation}
\label{Eq:LambdaInnerProduct.4.2.01}
\begin{array}{l} 
\displaystyle
\bfR\times\bfxi\cdot\hat\bfk\,=\,\hat{\bfk}\times \bfR\cdot\bfxi\,,
\end{array}
\end{equation}
which, once substituted into Equation~(\ref{Eq:LambdaInnerProduct.4.2}), gives
\begin{equation}
\label{Eq:LambdaInnerProduct.4.1.1}
\begin{array}{l} 
\displaystyle
\int\,\dd^2r_2\,
\exp(-\mathrm{i} u \bfr\times\bfr_2)\,
\Phi^*_{j',m'}(\bfr+\bfr_2)\,\Phi_{j,m}(\bfr_2)\,=\,\\
\\
\displaystyle
\,=\,(-1)^{2m}\,
\dfrac{\pi^2}{4\pi^2}\,\int\,\dd^2\bfxi\,
\exp\left(-\mathrm{i} \hat{\bfk}\times \bfR \cdot\bfxi\right)\,
\bar\Phi^*_{j,-m}\left(\bfR\,-\,\dfrac\bfxi 2\right)\,
\bar\Phi_{j',-m'}\left(\bfR\,+\,\dfrac\bfxi 2\right)\,=\,\\
\\
\displaystyle
\,=\,(-1)^{2m}\,\pi^2\,
\mathcal{W}\{\bar\Phi_{j,-m},\,\bar\Phi_{j',-m'}\}(\bfR,\,\hat{\bfk}\times \bfR)\,,
\end{array}
\end{equation}
where the symbol $\mathcal{W}\{\psi,\,\phi\}$ denotes the EWDF of the pair $\{\psi,\,\phi\}$, according to
the definition given in Ref.~\cite{Valkenburgh/2008}. 
To help our readers, 
the main definitions and properties of the EWDF have been briefly summarized in Appendix~\ref{Sec:WDFLGBis}, 
which also recalls the tight relationship between LG modes and  factorized HG modes through the Wigner distribution function (WDF henceforth)  and EWDF operators.
In particular, from~Equation~(\ref{Eq:WDFLGBis.4}), we~have
\begin{equation}
\label{Eq:LambdaInnerProduct.4.1.1.0.1}
\left\{
\begin{array}{l} 
\displaystyle
\bar\Phi_{j,-m}(\bfr)\,=\,(-1)^{j-|m|}\,\tilde {W}\{h_{j-m,j+m}\}(\bfr)\,=\,(-1)^{\min\{n,\ell\}}\,\tilde {W}\{h_{\ell,n}\}(\bfr),\\
\\
\bar\Phi_{j',-m'}(\bfr)\,=\,(-1)^{j'-|m'|}\,\tilde {W}\{h_{j'-m',j'+m'}\}(\bfr)\,=\,(-1)^{\min\{n',\ell'\}}\,\tilde {W}\{h_{\ell',n'}\}(\bfr)\,,
\end{array}
\right.
\end{equation}
where  the function  $\tilde {W}\{h_{h,k}\}(\bfr)$ is defined in Appendix~\ref{Sec:WDFLGBis}.
In the same appendix, the~{VanValkenburgh} 
 theorem has also been recalled (see Equation~(\ref{Eq:VVTheorem.2})); once applied to the present case, it gives
\begin{equation}
\label{Eq:LambdaInnerProduct.4.1.2Alternative}
\begin{array}{l}
\displaystyle
\mathcal{W}\{\bar\Phi_{j,-m},\,\bar\Phi_{j',-m'}\}({\bfR},\,\hat{\bfk}\times \bfR)\,=\,
\dfrac {(-1)^{\min\{n,\ell\}+\min\{n',\ell'\}}}\pi\,\tilde {W}\{h_{\ell,\ell'}\}\left({2X},\,{-2Y}\right)
\tilde {W}\{h_{n,n'}\}\left(0,\,0\right)\,.
\end{array}
\end{equation}
Here, for~the sake of comodity, we came back again to the pairs $(n,\ell)$ and $(n',\ell')$ 
in place of $(j,m)$ and $(j',m')$, respectively, with~$(X,Y)$ being the Cartesian representation of 
the transverse vector $\bfR$.
Now, as~far as the factor $\tilde {W}\{h_{n,n'}\}\left(0,\,0\right)$ is concerned, from~Equation~(\ref{Eq:WDFLGBis.4.1.1}) it is not difficult to obtain
\begin{equation}
\label{Eq:WDFLGBis.4.1.1Copia.2}
\begin{array}{l}
\displaystyle
\tilde W\{h_{n,n'}\}(0,0)\,=\,(-1)^{\min\{n,n'\}}
\bar\Phi_{\frac{n+n'}2,\frac{n-n'}2}(\boldsymbol{0})\,=\,\dfrac {(-1)^{n}}{\sqrt\pi}\,\delta_{n,n'}\,,
\end{array}
\end{equation}
where, in~the last step, use has been made of Equation~(\ref{Eq:ModalExpansionUniformTGSM.3}), evaluated at $u=1$. 
Equation~(\ref{Eq:WDFLGBis.4.1.1Copia.2}) is an important result. It implies that the matrix elements of 
$\hat W_u$ are necessarily ``diagonal'' with respect to the index pair $(n,n')$, 
{i.e.}
,
\begin{equation}
\label{Eq:WDFLGBis.4.1.1Copia.2.1}
\begin{array}{l}
\displaystyle
n \ne n' \Longrightarrow \bra n',\ell' |\hat W_u |n,\ell \ket \,=\,0
\end{array}
\end{equation}
Equation~(\ref{Eq:WDFLGBis.4.1.1Copia.2.1}) thus implies 
that the eigenket expansion of the state $\hat W_u|n,\ell\ket$ 
will involve only kets of the form $|n,\ell'\ket$. In~other words, the~double series expansion in Equation~(\ref{Eq:COSD.4.0.0.1.2.0.1})
reduces to a single series—a~nice mathematical surprise.
Generally speaking, we have
\begin{equation}
\label{Eq:COSD.4.0.0.1.2.0.1General}
\begin{array}{l}
\displaystyle
\hat W_u |n,\ell\ket\,=\,\sum^\infty_{\ell'=0}\,
\bra n,\ell'|\hat W_u |n,\ell\ket\,|n,\ell'\ket\,,\qquad\qquad \forall n,\ell\in\mathbb{N}.
\end{array}
\end{equation}
Moreover, we would also expect that, if~$\mu(-\bfr)=\mu(\bfr)$, then $\ell'$ must have 
the same parity as $\ell$. To~prove this, it is sufficient to note that the factor $\tilde W\{h_{\ell,\ell'}\}(2X,-2Y)$
turns out to be
\begin{equation}
\label{Eq:WDFLGBis.4.1.1Copia.3}
\begin{array}{l}
\displaystyle
\tilde W\{h_{\ell,\ell'}\}(2X,-2Y)\,=\,(-1)^{\min\{\ell,\ell'\}}
\bar\Phi_{\frac{\ell+\ell'}2,\frac{\ell-\ell'}2}(2X,-2Y)\,=\,
(-1)^{\min\{\ell,\ell'\}}
\bar\Phi^*_{\frac{\ell+\ell'}2,\frac{\ell-\ell'}2}(2\bfR)\,,
\end{array}
\end{equation}
where the second row of Equation~(\ref{Eq:LambdaInnerProduct.0.1}) has been used. 
Then, ~substituting from \mbox{Equations~(\ref{Eq:WDFLGBis.4.1.1Copia.2}) and~(\ref{Eq:WDFLGBis.4.1.1Copia.3})}
into Equation~(\ref{Eq:LambdaInnerProduct.4.1.2Alternative}), after~simple algebra we have
\begin{equation}
\label{Eq:WDFLGBis.4.1.1Copia.4}
\begin{array}{l}
\displaystyle
\mathcal{W}\{\bar\Phi_{j,-m},\,\bar\Phi_{j',-m'}\}({\bfR},\,\hat{\bfk}\times \bfR)\,=\,
\dfrac {(-1)^{n+\min\{\ell,\ell'\}+\min\{n,\ell\}+\min\{n',\ell'\}}}{\pi\sqrt\pi}\,
\bar\Phi^*_{\frac{\ell+\ell'}2,\frac{\ell-\ell'}2}(2\bfR)\,\delta_{n,n'}\,=\,\\
\\
\displaystyle
\,=\,\dfrac {(-1)^{n+\min\{\ell,\ell'\}+\min\{n,\ell\}+\min\{n',\ell'\}}}{\pi\sqrt{u\pi}}\,\Phi^*_{\frac{\ell+\ell'}2,\frac{\ell-\ell'}2}(\bfr)\,\delta_{n,n'}\,,
\end{array}
\end{equation}
where, in the last step, use has been made of $2\bfR=\bfr\sqrt u$. 
Finally,~substituting from Equation~(\ref{Eq:WDFLGBis.4.1.1Copia.4}) into 
Equation~(\ref{Eq:LambdaInnerProduct.4.1.1}) and taking into account that $(-1)^{2m}=(-1)^{n-\ell}$, we~have 
\begin{equation}
\label{Eq:LambdaInnerProductCopia}
\begin{array}{l} 
\displaystyle
\langle n',\ell' | \hat W_u |n,\ell\rangle\,=\,
(-1)^{
\ell+\min\{\ell,\ell'\}+\min\{n,\ell\}+\min\{n,\ell'\}}\,
\delta_{n,n'}\,
\sqrt{\dfrac\pi u}\,\int_{\mathbb{R}^2}\,
\dd^2 r\,
\mu(\bfr)\,
\Phi^*_{\frac{\ell+\ell'}2,\frac{\ell-\ell'}2}(\bfr)
\end{array}
\end{equation}
%
which represents the most important technical result of the present paper.
An immediate consequence of Equation~(\ref{Eq:LambdaInnerProductCopia}) is that,
under the hypothesis $\mu(-\bfr)=\mu(\bfr)$, it turns out to be
\begin{equation}
\label{Eq:WuMatrixElementsSymmetry}
\begin{array}{l} 
\displaystyle
\langle n,\ell' | \hat W_u |n,\ell\rangle\,\equiv\,0\,,
\end{array}
\end{equation}
for {\em any} pair $(\ell,\ell')$ such that $\ell-\ell'$ is an {\em odd integer}, as~shown in Appendix.~\ref{App:Symmetry}. 
In other words, the~orthonormal expansion of $\hat W_u|n,\ell\ket$ in Equation~(\ref{Eq:COSD.4.0.0.1.2.0.1General})
must involve only kets belonging to the subspace $\mathbb{V}_\pm$ containing $|n,\ell\ket$, as~it was conjectured~before.

{

\section{Twisting Schell-Model Sources with Radial Degrees of~Coherence}
\label{Sec:ProvingRadialMu}

{Equipped} 
 with the results found in the previous sections, it is now possible to provide 
{a} 
 rigorous 
proof of the results found in Ref.~\cite{Borghi/Gori/Santarsiero/Guattari/2015}.
To this end, we notice that, if~$\mu$ depends only on the radial variable $r$ and not on the azimuthal one $\varphi$, 
the symmetry condition of Equation~(\ref{Eq:MiuSymmetry}) is still verified. Moreover, the~integral in 
Equation~(\ref{Eq:LambdaInnerProductCopia}) factorizes as follows:
\begin{equation}
\label{Eq:LambdaInnerProductCopiaIntegralRadial}
\begin{array}{l} 
\displaystyle
\int_{\mathbb{R}^2}\,
\dd^2 r\,
\mu(\bfr)\,
\Phi^*_{\frac{\ell+\ell'}2,\frac{\ell-\ell'}2}(\bfr)\,\propto\,
\int_0^\infty\,\dd r\,f(r)\,\int_0^{2\pi}\,\dd\varphi \exp(-\ii (\ell-\ell') \varphi)\,,
\end{array}
\end{equation}
where, after~taking Equation~(\ref{Eq:ModalExpansionUniformTGSM.3}) into account, the~function $f(r)$ contains the product of sole radial factors, like 
$\mu(r)$, $r^{\ell+\ell'+1}$, and~the elegant Laguerre–Gauss function $\mathcal{L}^{\alpha}_k(ur^2)$
with indices $\alpha=|\ell-\ell'|$ and $k=\ell+\ell'-|\ell-\ell'|$. 
However, since the $\varphi$-integral turns out to be proportional to
$\delta_{\ell,\ell'}$, from~Equation~(\ref{Eq:LambdaInnerProductCopia}) it follows:
\begin{equation}
\label{Eq:LambdaInnerProductCopiaRadial}
\begin{array}{l} 
\displaystyle
\langle n,\ell' | \hat W_u |n,\ell\rangle\,=\,
\delta_{\ell,\ell'}\,
\sqrt{\dfrac\pi u}\,
\int_{\mathbb{R}^2}\,
\dd^2 r\,
\mu(\bfr)\,
\Phi^*_{\ell,0}(\bfr)\,, 
\end{array}
\end{equation}
which, after~taking Equation~(\ref{Eq:COSD.4.0.0.1.2.0.1General}) into account, implies that $|n,\ell\ket$ is an eigenket of $\hat W_u$, 
with eigenvalue given by
\begin{equation}
\label{Eq:LambdaInnerProductCopiaRadial.2}
\begin{array}{l} 
\displaystyle
\sqrt{\dfrac\pi u}\,
\int_{\mathbb{R}^2}\,
\dd^2 r\,
\mu(\bfr)\,
\Phi^*_{\ell,0}(\bfr)\,=\,
2\pi\,\int_0^\infty\,\dd r\,r\,\mu(r)\,\mathcal{L}_\ell(ur^2)\,,\qquad \ell=0,1,2,\ldots\,,
\end{array}
\end{equation}
where, for simplicity, $\mu(\bfr)=\mu(r)$ has been assumed.
Equation~(\ref{Eq:LambdaInnerProductCopiaRadial.2}) is consistent with the results presented
in Ref.~\cite{Borghi/Gori/Santarsiero/Guattari/2015} and gives them the promised mathematical rigor.
It would then be possible, differently from $\hat T_u$, 
to make $\hat W_u \ge 0$ by searching, for~a given radial spectral degree of coherence $\mu(r)$, those values of $u$ such that the infinite system of~inequalities 
\begin{equation}
\label{Eq:LambdaInnerProductCopiaRadial.2.1}
\begin{array}{l} 
\displaystyle
\int_0^\infty\,\dd r\,r\,\mu(r)\mathcal{L}_\ell(ur^2)\,\ge\,0\,,\qquad\qquad \ell=0,1,2,\ldots\,,
\end{array}
\end{equation}
be satisfied. 

It is now rather natural to explore the spectrum of $\hat W_u$ once the radial character of $\mu(\bfr)$ is removed. In~the next section, the~results proved in Sections~\ref{Sec:COSD} and~\ref{Sec:Integral} will be applied to a single, but~fundamental example of a non-radial degree of coherence. 
}

\section{On the Twistability of Anisotropic Gaussian Schell-Model~Sources}
\label{Subsec:ATGS}

In a seminal 1998 paper, Simon and Mukunda explored the structure of the most general class of paraxial 
shape-invariant partially coherent Gaussian Schell-model beams~\cite{Simon/Mukunda/1998b}. Inspired by
the achievements obtained by Simon and Mukunda, in~the present section we shall deal with the spectral degree of coherence
already considered in Section~\ref{Sec:COSD}, which is defined by
\begin{equation}
\label{Eq:ATGS.1}
\begin{array}{l} 
\displaystyle
\mu(\bfr)\,=\,\exp\left(-{x^2}-\chi\,y^2\right)\,,\qquad\qquad \chi\ge 1\,.
\end{array}
\end{equation}
Using group theory tools, Simon and Mukunda proved that, in~order for a twisted {\em anisotropic} Gaussian source obtained from the degree of coherence of  Equation~(\ref{Eq:ATGS.1}) to be {bonafide}
, the~following condition (translated into our notation) must be fulfilled:
\begin{equation}
\label{Eq:ATGS.1.1}
\begin{array}{l} 
\displaystyle
u^2  \le 4\chi \,.
\end{array}
\end{equation}
The scope of the present section is to explore the role of the Simon–Mukunda condition~(\ref{Eq:ATGS.1.1}) inside the 
spectrum of the corresponding twisted operator $\hat W_u$. To~this end, all matrix elements of the operator 
$\hat W_u$ obtained by twisting the degree of coherence in 
Equation~(\ref{Eq:ATGS.1}) will be evaluated in closed form.  
First of all, since the matrix elements $\bra n,\ell'|\hat W_u|n,\ell\ket$ turn out to be real for 
any triplet $(n,\ell,\ell')$, in~the following it will supposed that $\ell' \ge \ell$. Moreover, all non-vanishing matrix elements
correspond to indices $\ell$ and $\ell'$ having the same parity. Accordingly, from~
Equation~(\ref{Eq:LambdaInnerProductCopia}), we have
\begin{equation}
\label{Eq:LambdaInnerProductCopiaA}
\begin{array}{l} 
\displaystyle
\langle n,\ell+2M | \hat W_u |n,\ell\rangle\,=\,\langle n,\ell | \hat W_u |n,\ell+2M\rangle\,=\,\\
\\
\,=\,
\displaystyle
(-1)^{\min\{n,\ell\}+\min\{n,\ell+2M\}}\,\sqrt{\dfrac\pi u}\,\int_{\mathbb{R}^2}\,
\dd^2 r\,
\mu(\bfr)\,
\Phi^*_{\ell+M,M}(\bfr)\,,
\end{array}
\end{equation}
where the integral can be explicited, as~shown in Appendix~\ref{App:MatrixElements}, as~follows:
\begin{equation}
\label{Eq:ATGS.1.2}
\begin{array}{l} 
\displaystyle
\sqrt{\dfrac\pi u}\int_{\mathbb{R}^2}\,
\dd^2r\,\mu(\bfr)\,\Phi^*_{\ell+M,M}\,=\, 2\pi\,u^M\,
\sqrt{\dfrac{\ell!}{(\ell+2M)!}}\,
\\
\\
\displaystyle
\times\,\int_0^\infty\,\dd r\,
r^{2M+1}\,L^{2M}_{\ell}(ur^2)\,I_M\left(\dfrac{r^2}2(\chi-1)\right)\,
\exp\left(-\dfrac{r^2}2 (\chi+u+1)\right)\,,
\end{array}
\end{equation}
%
where the symbol $I_n(\cdot)$ denotes the $n$th-order modified Bessel function of the first kind
~\cite{DLMF}.
The integral in Equation~(\ref{Eq:ATGS.1.2}) can be expressed in closed-form terms, starting from a notable 
expression recently found by Yuri Aleksandrovich Brychkov~\cite{Brychkov/2014} involving 
an important class of bivariate hypergeometric functions—the~so-called Appell \mbox{functions~\cite{DLMF}~(Ch. 16)}.
First of all,  after~the integration variable change $x=r^2/2$, Equation~(\ref{Eq:ATGS.1.2}) becomes
\begin{equation}
\label{Eq:ATGS.1.2.1}
\begin{array}{l} 
\displaystyle
\sqrt{\dfrac \pi u}\,
\int_{\mathbb{R}^2}\,
\dd^2r\,\mu(\bfr)\,
\Phi^*_{\ell+M,M}\,=\,2\pi\,(2u)^M\,
\sqrt{\dfrac{\ell!}{(\ell+2M)!}}\,\\
\\
\displaystyle
\times\,\int_0^\infty\,\dd x\,
x^{M}\,L^{2M}_{\ell}(2ux)\,I_M\left((\chi-1)x\right)\,
\exp\left(-(\chi+u+1)x\right)\,.
\end{array}
\end{equation}
Since all subsequent mathematical steps turned out to be not trivial, cumbersome, and~somewhat boring, 
they have been confined to Appendix~\ref{App:Brychkov}, where it is proved that for~\mbox{$a>b\ge 0$} and
for $n,m \in \mathbb{N}$,
\begin{equation}
\label{Eq:BrychkovNewFormulaSimplifiedGegenbauer}
\begin{array}{l}
\displaystyle
\int_0^\infty\,\dd x\,
x^{m}\,\exp(-a\,x)\,I_m(b\,x)\,L^{2m}_n(c\,x) \,=\,\\
\\
\displaystyle
\,=\,\left(\dfrac b2\right)^m\,
\dfrac{(2m+n)!}{(m+n)!}\,
\dfrac{(-bc/2)^n}{(a^2-b^2)^{n+m+1/2}}
C^{-m-n}_n\left(\dfrac{a^2-b^2-ac}{bc}\right),
\end{array}
\end{equation}
where the symbol $C^\lambda_n(\cdot)$ denotes the so-called ultraspherical, or~Gegenbauer polynomials~\cite{DLMF}.
Substituting 
Equation~(\ref{Eq:BrychkovNewFormulaSimplifiedGegenbauer}) into Equation~(\ref{Eq:LambdaInnerProductCopiaA}),
we have 
\begin{equation}
\label{Eq:ATGS.1.2.1}
\begin{array}{l} 
\displaystyle
\langle n,\ell+2M| \hat W_u |n,\ell\rangle \,=\, \langle n,\ell | \hat W_u |n,\ell+2M\rangle
\,=\,
\dfrac{2\pi\,(-1)^{\min\{n,\ell\}+\min\{n,\ell+2M\}}}{\sqrt{(u+2)(u+2\chi)}}\\
\\
\displaystyle
\times
\dfrac{\sqrt{\ell!\,(\ell+2M)!}}{(\ell+M)!}\,
\left[\dfrac{u(\chi-1)}{(u+2)(u+2\chi)}\right]^{l+M}\,
C^{-\ell-M}_\ell\left(\dfrac{u^2-4\chi}{2u(\chi-1)}\right).
\end{array}
\end{equation}
This equation, together with Equation~(\ref{Eq:LambdaInnerProductCopia}), constitutes another important result of the present paper,
since
it unveils the complete information about the coherence features of the twisted uniform source obtained 
from the anisotropic degree of coherence~(\ref{Eq:ATGS.1}). 
In particular, the~role played by inequality~(\ref{Eq:ATGS.1.1}) can be grasped directly from the matrix element arrangement. 
According to Simon and Mukunda and to what has been exposed here so far, we assert that under~the condition 
$u^2 \le 4\chi$, {\em any} twisted Schell-model source of the form
\begin{equation}
\label{Eq:ATGS.1.2.1.1}
\begin{array}{l} 
\displaystyle
W(\bfr_1,\bfr_2)\,=\,\tau(\bfr_1)\,\tau^*(\bfr_2)\,\exp(-(x_1-x_2)^2-\chi\,(y_1-y_2)^2)\,\exp(-\ii u \bfr_1\times\bfr_2)\,,
\end{array}
\end{equation}
turns out to be {bonafide}
, irrespective the choice of the amplitude filter $\tau(\bfr)$.
That \mbox{Equation~(\ref{Eq:ATGS.1.1})} represents a {\em necessary} condition can be proved by imposing 
that all ``diagonal'' matrix elements of the uniform operator 
obtained from Equation~(\ref{Eq:ATGS.1.2.1.1}), assuming $\tau\equiv 1$, i.e.,~$\bra n,\ell|\hat W_u|n,\ell\ket$, are non-negative
$\forall n,\ell \in \mathbb{N}^2$.
Proving such a statement can be achieved via 
Equation~(\ref{Eq:ATGS.1.2.1}) 
written for $M=0$, which gives
\begin{equation}
\label{Eq:ATGS.1.2.1.0.0.0.1}
\begin{array}{l} 
\displaystyle
\langle n,\ell | \hat W_u |n,\ell\rangle\,=\,
\dfrac{2\pi}{\sqrt{(u+2)(u+2\chi)}}
\left[\dfrac{u(\chi-1)}{(u+2)(u+2\chi)}\right]^{\ell}\,
C^{-\ell}_\ell\left(\dfrac{u^2-4\chi}{2u(\chi-1)}\right)\,.
\end{array}
\end{equation}
The first two factors are inherently non-negative, due to the hypothesis $\chi\ge 1$. To~find the conditions under which the
ultraspherical polynomial is also non-negative ($\forall \ell\ge 0$), a~recent theorem proved by Driver and Duren~\cite{Driver/Duren/2000,Driver/Duren/2001} states that all zeros of the Gegenbauer polynomials $C^{\lambda}_\nu(x)$ are purely imaginary
(zero included), provided that $\lambda < 1-\nu$. This, together with the following elementary  asymptotics~\cite{DLMF} (formula 18.5.10):
\begin{equation}
\label{Eq:ATGS.1.2.1.0.0.0.1.1}
\begin{array}{l} 
\displaystyle
C^{-\ell}_\ell(x)\,\sim\, (-2x)^\ell,\qquad\qquad |x| \to \infty\,,
\end{array}
\end{equation}
confirms that, in~order for the matrix elements in Equation~(\ref{Eq:ATGS.1.2.1.0.0.0.1}) to be identically non-negative,
the argument of the ultraspherical polynomial must be non-positive $\forall \ell \ge 0$. 
{Thus, the Simon–Mukunda inequality in Equation~(\ref{Eq:ATGS.1.1}) is {\em necessary} for~$\hat W_u$ to be semidefinite~positive}.

What could be surprising, at~least in the present case, is that the same condition also turns out to be sufficient, as~proved 
in~\cite{Simon/Mukunda/1998b}, for the particular case of the anisotropic TGSM beams.  
Several numerical simulations, again carried out via {\em Mathematica}, have been performed to check 
such an intriguing result within the new framework of the matrix element representation of $\hat W_u$.
However, rather than annoy our readers with a dull description of such simulations, our attention will be focused
on a single case, in~which analytical results can be found.
Before doing this, it is worth summarizing the terms of the problem, in~consideration of what has been shown so far.
Due to the high spectral degeneration of the $\hat T_u$ operator, the~action of  $\hat W_u$ on a typical  $\hat T_u$ eigenstate
$|n,\ell\ket$ produces the state $\hat W_u |n,\ell\ket$ belonging to $\mathbb{V}_+$ ($\mathbb{V}_-$)
if $\ell$ is an even (odd) integer. Thanks to the ``diagonal'' character of the matrix elements 
(as expressed by Equation~(\ref{Eq:LambdaInnerProductCopia})), 
Equation~(\ref{Eq:COSD.4.0.0.1.2.0.1}) reduces to Equation~(\ref{Eq:COSD.4.0.0.1.2.0.1General}), which dramatically restricts 
the (infinite) set of the eigenstates involved in the representation of $\hat W_u |n,\ell\ket$. 
The complete knowledge of the symmetric matrix $\{\bra n,\ell'|\hat W_u |n,\ell\ket\}^\infty_{\ell,\ell'=0}$ obtained through 
Equation~(\ref{Eq:COSD.4.0.0.1.2.0.1}) would allow, in~principle, ~solving such a degenerate eigenvalue~problem.

As anticipated, we shall limit ourselves to the so-called {\em saturated case}, namely $u^2=4\chi$, which corresponds 
to the maximum twist strength admissible for a given $\chi$.  
Assuming $u=2\sqrt\chi$,
the matrix element~(\ref{Eq:ATGS.1.2.1}) can be rearranged as follows: 

\begin{equation}
\label{Eq:ATGS.1.2.1.New}
\begin{array}{l} 
\displaystyle
\langle n,\ell+2M| \hat W_u |n,\ell\rangle \,=\,
\langle n,\ell | \hat W_u |n,\ell+2M\rangle\,=\,\dfrac{\pi}{\chi^{1/4}\,(1+\sqrt{\chi})}\\
\\
\displaystyle
\times\,
(-1)^{\min\{n,\ell\}+\min\{n,\ell+2M\}}\,
\dfrac{\sqrt{\ell!\,(\ell+2M)!}}{(\ell+M)!}\,
\xi^{l+M}\,
C^{-\ell-M}_\ell(0)\,,
\end{array}
\end{equation}
where the real parameter $\xi$, defined as
\begin{equation}
\label{Eq:ATGS.1.2.1.New.1}
\begin{array}{l} 
\displaystyle
\xi\,=\,\dfrac 12\,\dfrac{\sqrt\chi-1}{\sqrt\chi+1}\,,
\end{array}
\end{equation}
runs within the interval $[0,\frac 12[$.
The structure of Equation~(\ref{Eq:ATGS.1.2.1.New}) is particularly interesting. First of all, due to the fact that the 
Gegenbauer polynomial $C^{\lambda}_\ell(x)$ has the same parity of the index $\ell$, all elements corresponding
to odd values of $\ell$ are necessarily null. 
More importantly, after~re-introducing the index $\ell'=\ell+2M$, 
the matrix element in~(\ref{Eq:ATGS.1.2.1.New}) can be {\em factorized} as follows:
\begin{equation}
\label{Eq:ATGS.1.2.1.New.0}
\begin{array}{l} 
\displaystyle
\langle n,\ell| \hat W_u |n,\ell'\rangle \,=\,
\phi_{n,\ell}\,\phi_{n,\ell'}\,,
\end{array}
\end{equation}
where
\begin{equation}
\label{Eq:ATGS.1.2.1.New.1.1}
\begin{array}{l} 
\displaystyle
\phi_{n,\ell}\,=\,
\sqrt{\dfrac{\pi}{\chi^{1/4}\,(1+\sqrt{\chi})}}\,\times\,
\left\{
\begin{array}{cc} 
(-1)^{\min\{n,\ell\}}\,\dfrac{\sqrt{\ell !}}{(\ell/2)!}\,\xi^{\ell/2}\,,& \ell\,\,\,\mathrm{even}\,,\\
& \\
0\,,&  \ell\,\,\,\mathrm{odd}\,,
\end{array}
\right.
\end{array}
\end{equation}
and use has been made of the fact that~\cite{DLMF} (Table 18.6.1),
\begin{equation}
\label{Eq:ATGS.1.2.1.New.1.1.1}
\begin{array}{l} 
\displaystyle
C^{(\lambda)}_{2n}(0)\,=\,(-1)^n\,\dfrac{(\lambda)_n}{n!}\,.
\end{array}
\end{equation}
From Equation~(\ref{Eq:ATGS.1.2.1.New.0}), it follows that the spectrum of the matrix
$\{\bra n,\ell'|\hat W_u |n,\ell\ket\}^\infty_{\ell,\ell'=0}$ turns out to be of the form
$\{\Lambda,\,0,\,0,\,0,\,\ldots\}$, with~$\Lambda$ being its trace.
After \mbox{using~\cite{PrudnikovI} (formula 5.2.13.1),} we have
\begin{equation}
\label{Eq:ATGS.1.2.1.New.1.1.2}
\begin{array}{l} 
\displaystyle
\Lambda\,=\,
\sum_{\ell=0}^\infty\,
\,\bra n,\ell|\hat W_u |n,\ell\ket\,=\,
\dfrac{\pi}{\chi^{1/4}\,(1+\sqrt{\chi})}\,
\sum_{k=0}^\infty\,\dfrac{(2k)!}{k!^2}\,\xi^{2k}\,=\,
\dfrac\pi{2\sqrt\chi}\,,
\end{array}
\end{equation}
where, in the last step, use has been made of Equation~(\ref{Eq:ATGS.1.2.1.New.1}).
Since the value of $\Lambda$ does not depend on the index $n$, we conclude that the complete spectrum
of the twisted operator $\hat W_u$ consists of an infinite list of infinite lists, i.e.,
\begin{equation}
\label{Eq:ATGS.1.2.1.New.1.1.3.4}
\begin{array}{l} 
\displaystyle
\left\{
\left\{\dfrac\pi{2\sqrt\chi},\,0,\,0,\,0,\,\ldots\right\},\,
\left\{\dfrac\pi{2\sqrt\chi},\,0,\,0,\,0,\,\ldots\right\},\,
\left\{\dfrac\pi{2\sqrt\chi},\,0,\,0,\,0,\,\ldots\right\},\,
\ldots
\right\}\,,
\end{array}
\end{equation}
which confirms the positive semidefiniteness of $\hat W_u$.

\section{Conclusions}
\label{Sec:Conclusions}

More than thirty years have passed since the birth of TGSM beams, which marked the introduction of {\em twist} 
into the scenario of classical coherence theory. Despite hundreds of works published on the subject, there remain  
still unexplored areas that deserve to be studied. {Among them, the~problem of identifying 
the conditions under which untwisted {bonafide} 
 CSDs can made twistable simply by imposing them a twist term 
$\exp(-\ii  u \bfr_1\times \bfr_2)$ seems to be far from being resolved. 
In the present paper, a~possible strategy has been outlined to solve, at~least in principle, such an ambitious technical problem
for the class of Schell-model sources.}
In particular, by~employing the operatorial approach introduced in~\cite{Simon/Mukunda/1998}, a~complete characterization of a twisted,
uniform operator $\hat W_u$ has been analytically performed in terms of the eigenstates of the sole twist operator $\hat T_u$,  
thanks to the important results of Ref.~\cite{Valkenburgh/2008}. Such a choice has been driven by the fact that, for~the class 
of {\em real} symmetric functions $\mu(\bfr)=\mu(-\bfr)$ considered here, the~operators $\hat W_u$ and $\hat T_u$ must commute. 
{The algorithm developed here has also allowed the results of Ref.~\cite{Borghi/Gori/Santarsiero/Guattari/2015}, where the subclass of {\em radial} functions $\mu(|\bfr|)$ was studied, to~be given a solid mathematical justification. 
Finally, to~illustrate a nontrivial application, the~class of the twisted sources obtained by non-radial Gaussian degrees of coherence has also been analyzed,~considering the numerical 
limitation of the twist strength found in~\cite{Simon/Mukunda/1998}.}

\acknowledgments{I wish to thank Turi Maria Spinozzi for his useful comments and~help.
{I am also grateful to  Ari Friberg, Andreas Norrman and Tero Set\"al\"a for inviting me, last June, at the 
3rd Conference on Coherence and Random Polarization in Joensuu, Finland,
where the main idea about this paper started.}\\
{To Marina Ettorri (1970 - 2024), \emph{in Memoriam}.}
}


\vspace{6pt} 


\appendix


\section{Proof of Eq.~(\ref{Eq:Notations.5.1})}
\label{App:Commutation}

In order to find the conditions under which Eq.~(\ref{Eq:Notations.5.1}) holds, we have to impose that, for
any $|\psi\ket \in \mathcal{H}$, the following equation must hold:
\begin{equation}
\label{Eq:App:Commutation.1}
\begin{array}{l}
\displaystyle
\hat W_u\hat T_u|\psi\ket=\,\hat T_u\hat W_u|\psi\ket,\qquad\qquad \forall |\psi\ket \in \mathcal{H}.
\end{array}
\end{equation}
The matrix elements of the operator $\hat W_u\hat T_u$ are given, in the position representation, by
\begin{equation}
\label{Eq:App:Commutation.1.1}
\begin{array}{l}
\displaystyle
\bra \bfr_1 |\hat W_u\hat T_u|\bfr_2\ket\,=\,
\int_{\mathbb{R}^2}\,\dd^2 \rho\,
\bra \bfr_1 |\hat W_u|\bfrho\ket\bra \bfrho \hat T_u|\bfr_2\ket\,=\,
\int_{\mathbb{R}^2}\,\dd^2 \rho\,
 W_u(\bfr_1,\bfrho)\,T_u(\bfrho,\bfr_2)\,=\,\\
\\
\,=\,
\displaystyle
\int_{\mathbb{R}^2}\,\dd^2 \rho\,
 \mu(\bfr_1-\bfrho)\,\exp(-\ii u (\bfr_1-\bfr_2)\times\bfrho)\,=\,
\displaystyle
\int_{\mathbb{R}^2}\,\dd^2 \rho\,
 \mu(\bfxi)\,\exp(-\ii u \bfxi\times (\bfr_1-\bfr_2))\,=\,\\
\\
\,=\,
\displaystyle
\exp(-\ii u \bfr_1\times\bfr_2)
\int_{\mathbb{R}^2}\,\dd^2 \bfxi\,
\mu(\bfxi)\,\exp(-\ii u \bfxi\times (\bfr_1-\bfr_2))\,,
\end{array}
\end{equation}
where the new integration variable $\bfxi=\bfr_1-\bfrho$ has been introduced. 

Similarly,  for the product $\hat T_u\hat W_u$ we have 
\begin{equation}
\label{Eq:App:Commutation.1.2}
\begin{array}{l}
\displaystyle
\bra \bfr_1 |\hat T_u\hat W_u|\bfr_2\ket\,=\,
\int_{\mathbb{R}^2}\,\dd^2 \rho\,
 T_u(\bfr_1,\bfrho)\,W_u(\bfrho,\bfr_2)\,=\,\\
\\
\displaystyle
\,=\,\int_{\mathbb{R}^2}\,\dd^2 \rho\,
\exp(-\ii u \bfr_1\times\bfrho)\,\mu(\bfrho-\bfr_2)\,\exp(-\ii u \bfrho\times\bfr_2)\,=\,\\
\\
\,=\,
\displaystyle
\exp(-\ii u \bfr_1\times\bfr_2)
\int_{\mathbb{R}^2}\,\dd^2 \bfxi\,
\mu(-\bfxi)\,\exp(-\ii u \bfxi\times (\bfr_1-\bfr_2))\,,
\end{array}
\end{equation}
where now we set $-\bfxi=\bfrho-\bfr_2$. On comparing Eq.~(\ref{Eq:App:Commutation.1.1})
with Eq.~(\ref{Eq:App:Commutation.1.2}), we conclude that the commutator $[\hat W_u,\hat T_u]$ vanishes iff $\mu(\bfxi)\equiv\mu(-\bfxi)$. 	

\section{Proof of Eq.~(\ref{Eq:WuMatrixElementsSymmetry})}
\label{App:Symmetry}

Consider again the quantity $\langle n,\ell' | \hat W_u |n,\ell\rangle$ given by Eq.~(\ref{Eq:LambdaInnerProductCopia}),
\begin{equation}
\label{Eq:App:Symmetry.1}
\begin{array}{l} 
\displaystyle
\langle n,\ell' | \hat W_u |n,\ell\rangle\,=\,
\displaystyle
\sqrt{\dfrac\pi u}\,
(-1)^{\ell+\min\{\ell,\ell'\}+\min\{n,\ell'\}+\min\{n,\ell\}}\,
\int_{\mathbb{R}^2}\,
\dd^2 r\,
\mu(\bfr)\,
\Phi^*_{\frac{\ell'+\ell}2,\frac{\ell-\ell'}2}(\bfr),
\end{array}
\end{equation}
and introduce two auxiliary variables, say $J=\frac{\ell+\ell'}2 \ge 0$ and $M=\frac{\ell-\ell'}2$.

Suppose now that  $2M$ be an {\em odd} integer number. 
Since by hypothesis $\mu(-\bfr)=\mu(\bfr)\in\mathbb{R}$, on using the second of Eq.~(\ref{Eq:LambdaInnerProduct.0.1}),
the integral into Eq.~(\ref{Eq:App:Symmetry.1}) becomes
\begin{equation}
\label{Eq:App:Symmetry.1.1}
\begin{array}{l} 
\displaystyle
\int_{\mathbb{R}^2}\,
\dd^2 r\,
\mu(\bfr)\,
\Phi^*_{J,M}(\bfr)\,=\,
\displaystyle
\int_{\mathbb{R}^2}\,
\dd^2 r\,
\mu(-\bfr)\,
\Phi^*_{J,M}(-\bfr)\,=\,
\displaystyle
\int_{\mathbb{R}^2}\,
\dd^2 r\,
\mu(\bfr)\,
\Phi^*_{J,M}(-\bfr)\,=\,\\
\\
\,=\,
(-1)^{2M}\,\displaystyle
\int_{\mathbb{R}^2}\,
\dd^2 r\,
\mu(\bfr)\,
\Phi^*_{J,M}(\bfr)\,=\,
-\int_{\mathbb{R}^2}\,
\dd^2 r\,
\mu(\bfr)\,
\Phi^*_{J,M}(\bfr)\,,
\end{array}
\end{equation}
which implies that the integral is null and, consequently, also the matrix element. 
Accordingly,  $\langle n,\ell' | \hat W_u |n,\ell\rangle\,=\,0$ for any odd $\ell-\ell'$. 


\section{On the VanValkenburgh extended Wigner distribution function 
}
\label{Sec:WDFLGBis}

We start from the definition of Wigner distribution function (WDF henceforth)
of a given coherent wavefield $\psi(\bfr)$, namely~\cite{Simon/Agarwal/2000}
\begin{equation}
\label{Eq:WDFLGBis.1}
\begin{array}{l}
\displaystyle
\mathcal{W}\{\psi\}(\bfrho,\bfp)\,=\,\dfrac 1{4\pi^2}\,\int_{\mathbb{R}^2}\,\dd^2\xi\,
\exp(-\ii \bfxi\cdot\bfp)\,
\psi^*\left(\bfrho\,-\,\dfrac{\bfxi}2\right)
\psi\left(\bfrho\,+\,\dfrac{\bfxi}2\right)\,.
\end{array}
\end{equation}
In 2008, VanValkenburgh proposed the following natural generalization of the WDF~\cite{Valkenburgh/2008}:
\begin{equation}
\label{Eq:WDFLGBis.2}
\begin{array}{l}
\displaystyle
\mathcal{W}\{\psi,\phi\}(\bfrho,\bfp)\,=\,\dfrac 1{4\pi^2}\,\int_{\mathbb{R}^2}\,\dd^2\xi\,\exp(-\ii \bfxi\cdot\bfp)\,
\psi^*\left(\bfrho\,-\,\dfrac{\bfxi}2\right)
\phi\left(\bfrho\,+\,\dfrac{\bfxi}2\right)\,, 
\end{array}
\end{equation}
where now $\psi(\bfr)$ e $\phi(\bfr)$ denote {\em arbitrary} complex wavefields.
In particular, if $\psi \equiv \phi$, then the EWDF defined into Eq.~(\ref{Eq:WDFLGBis.2}) must 
coincide with the ordinary WDF defined into Eq.~(\ref{Eq:WDFLGBis.1}), i.e., 
\begin{equation}
\label{Eq:WDFLGBis.2.1}
\begin{array}{l}
\displaystyle
\mathcal{W}\{\psi,\psi\}(\bfrho,\bfp)\,\equiv\,\mathcal{W}\{\psi\}(\bfrho,\bfp)\,.
\end{array}
\end{equation}
A quarter of century ago, Simon and Agarval first evaluated the WDF of LG modes~(\ref{Eq:ModalExpansionUniformTGSM.3}), namely~\cite{Simon/Agarwal/2000}
\begin{equation}
\label{Eq:WDFLGBis.3.1}
\begin{array}{l}
\displaystyle
\mathcal{W}\{\Phi_{j,m}\}(\bfrho,\bfp)\,=\,
\frac{(-1)^{2j}}{\pi^2}\,\mathcal{L}_{j+m}(Q_0+Q_2)\,\mathcal{L}_{j-m}(Q_0-Q_2)\,,
\end{array}
\end{equation}
where  $\mathcal{L}_n(x)=L_n(x) \exp(-x/2)$ and 
the quantities $Q_0$ and $Q_2$ were defined by
\begin{equation}
\label{Eq:WDFLGBis.3.2}
\left\{
\begin{array}{l}
\displaystyle
Q_0\,=\, {u}{\rho^2}+\dfrac{p^2}u\,,\\
\\
\displaystyle
Q_2\,=\,2\,\bfrho\times\bfp\,.
\end{array}
\right.
\end{equation}
Eigth years later, such a fundamental result was analyzed from a different point of view by VanValkenburgh~\cite{Valkenburgh/2008}, 
who proved that any LG mode $\Phi_{j,m}$ can be interpreted as an EWDF and, more importantly, conceived a
beautiful and simple analytical algorithm to express the EWDF of the product of two {\em arbitrarily chosen} LG modes, 
thus extending the achievement obtained in~\cite{Simon/Agarwal/2000}.

In order to ``tune'' the notations of the VanValkenburgh paper with those, slightly different,  
of the Simon/Agarwal paper, it will assumed $u=1$. In this way, it can be proved that~\cite{Valkenburgh/2008} 
\begin{equation}
\label{Eq:WDFLGBis.4}
\begin{array}{l}
\displaystyle
\bar\Phi_{j,m}(\bfr)\,=\,
(-1)^{j-|m|}\,\tilde{W}\{h_{j+m,j-m}\}(x,y)\,, 
\end{array}
\end{equation}
where $h_{n,\ell}(x,y)=h_n(x)\,h_\ell(y)$, with the so-called HG functions $h_n(x)$ 
being defined as
\begin{equation}
\label{Eq:WDFLGBis.3}
\begin{array}{l}
\displaystyle
h_n(x)\,=\,\pi^{-1/4} n!^{-1/2} 2^{-n/2}\,\exp\left(-\dfrac 12 x^2\right)\,H_n(x),\qquad n=0,1,2,\ldots
\end{array}
\end{equation}
As far as the symbol $\tilde{W}\{h_{n,\ell}\}(x,y)$ is concerned, again from Ref.~\cite{Valkenburgh/2008} we have
\begin{equation}
\label{Eq:WDFLGBis.3.1.1.1.1}
\begin{array}{l}
\displaystyle
\tilde {W}\{h_{n,\ell}\}(x,y)\,=\,
\dfrac 1{\sqrt{2\pi}}\,
\int_{\mathbb{R}}\,\dd\xi\,
\exp(-\ii \xi y)\,
h_{n,\ell}\left(\dfrac{x\,-\,\xi}{\sqrt 2},\,\dfrac{x\,+\,\xi}{\sqrt 2}\right)\,.
\end{array}
\end{equation}
In particular, Eq.~(\ref{Eq:WDFLGBis.4}) can be inverted as follows: 
\begin{equation}
\label{Eq:WDFLGBis.4.1.1}
\begin{array}{l}
\displaystyle
\tilde W\{h_{n,\ell}\}(x,y)\,=\,(-1)^{\min\{n,\ell\}}
\bar\Phi_{\frac{n+\ell}2,\frac{n-\ell}2}(\bfr)\,,
\end{array}
\end{equation}
which plays a role of pivotal importance for the scope of the present paper.

To illustrate the VanValkenburgh theorem, consider {\em his own} definition of EWDF, precisely~\cite{Valkenburgh/2008}
\begin{equation}
\label{Eq:WDFLGBis.5}
\begin{array}{l}
\displaystyle
{W}_2\{\psi,\phi\}(\bfrho,\bfp)\,=\,
\dfrac 1{2\pi}\,\int_{\mathbb{R}^2}\,\dd^2\xi\,\exp(-\ii \bfxi\cdot\bfp)\,
\psi^*\left(\dfrac{\bfrho\,-\,\bfxi}{\sqrt 2}\right)
\phi\left(\dfrac{\bfrho\,+\,\bfxi}{\sqrt 2}\right)\,,
\end{array}
\end{equation}
which, as said above,  slightly differs from the definition given into Eq.~(\ref{Eq:WDFLGBis.2}).
Nevertheless, it is trivial to prove that  
\begin{equation}
\label{Eq:WDFLGBis.6}
\begin{array}{l}
\displaystyle
\mathcal{W}\{\psi,\phi\}({\bfrho},\,{\bfp})\,=\,
\dfrac 1\pi\,{W}_2\{\psi,\phi\}\left(\bfrho\sqrt 2,\bfp\sqrt 2\right)\,.
\end{array}
\end{equation}
Now, in Ref.~\cite{Valkenburgh/2008} it has been proved that:
\begin{equation}
\label{Eq:VVTheorem}
\begin{array}{l}
\displaystyle
{W}_2\{\tilde {W}\{h_{n,\ell}\},\tilde {W}\{h_{n',\ell'}\}\}(\bfrho,\bfp)\,=\,
\tilde {W}\{h_{n,n'}\}\left(\dfrac{x\,+\,p_y}{\sqrt 2},\,\dfrac{p_x\,-\,y}{\sqrt 2}\right)
\tilde {W}\{h_{\ell,\ell'}\}\left(\dfrac{x\,-\,p_y}{\sqrt 2},\,\dfrac{p_x\,+\,y}{\sqrt 2}\right),
\end{array}
\end{equation}
which, on taking Eq.~(\ref{Eq:WDFLGBis.6}) into account, leads to
\begin{equation}
\label{Eq:VVTheorem.2}
\begin{array}{l}
\displaystyle
\mathcal{W}\{\tilde {W}\{h_{n,\ell}\},\tilde {W}\{h_{n',\ell'}\}\}({\bfrho},\,{\bfp})\,=\,
\dfrac 1\pi\,\tilde {W}\{h_{n,n'}\}\left({x\,+\,p_y},\,{p_x\,-\,y}\right)
\tilde {W}\{h_{\ell,\ell'}\}\left({x\,-\,p_y},\,{p_x\,+\,y}\right).
\end{array}
\end{equation}

\section{Proof of Eq.~(\ref{Eq:ATGS.1.2})}
\label{App:MatrixElements}

We start from Eq.~(\ref{Eq:ModalExpansionUniformTGSM.3}) written in polar coordinates $(r,\varphi)$,
\begin{equation}
\label{Eq:App:MatrixElements.1}
\begin{array}{l} 
\displaystyle
\sqrt{\dfrac \pi u}\int_{\mathbb{R}^2}\,
\dd^2 r\,
\mu(\bfr)\,
\Phi^*_{\ell,\ell+2M}(\bfr)\,=\,
u^M\,\sqrt{\dfrac{\ell!}{(\ell+2M)!}}
\int_0^\infty\,\dd r\,r^{2M+1}\,L^{2M}_{J-M}(ur^2)\,\exp\left(\dfrac{ur^2}2\right)\,\,\\
\\
\times
\displaystyle
\int_0^{2\pi}\,\dd \varphi\,
\exp\left[-r^2\,\left(\cos^2\varphi\,+\,\chi\,\sin^2\varphi\right)\right]\,\exp(-2\ii M \varphi)\,.
\end{array}
\end{equation}
The $\varphi$-integral can be evaluated elementarly, 
\begin{equation}
\label{Eq:App:MatrixElements.2}
\begin{array}{l} 
\displaystyle
\int_0^{2\pi}\,\dd \varphi\,
\exp\left[-r^2\,\left(\cos^2\varphi\,+\,\chi\,\sin^2\varphi\right)\right]\,\exp(-2\ii M \varphi)\,=\,\\
\\
\,=\,
\displaystyle
\exp\left[-\dfrac{r^2}2(\chi+1)\right]\,
\int_0^{2\pi}\,\dd \varphi\,
\exp\left[-\dfrac{r^2}2(\chi-1)\,\cos 2\varphi\right]\,
\,\exp(-2\ii M \varphi)\,=\,\\
\\
\,=\,
2\pi\,\exp\left[-\dfrac{r^2}2(\chi+1)\right]\,
I_M\left(\dfrac{r^2}2(\chi-1)\right)\,,
\end{array}
\end{equation}
where $I_n(\cdot)$ denotes the $n$th-order modified Bessel function of the first kind.
On substituting from Eq.~(\ref{Eq:App:MatrixElements.2}) into Eq.~(\ref{Eq:App:MatrixElements.1}),
after rearranging Eq.~(\ref{Eq:ATGS.1.2}) follows.


\section{Proof of Eq.~(\ref{Eq:BrychkovNewFormulaSimplifiedGegenbauer})}
\label{App:Brychkov}

The starting point is the following result by Brichkov~\cite[formula~14.2.18]{Brychkov/2014}: 
\begin{equation}
\label{Eq:BrychkovNewFormula}
\begin{array}{c}
\displaystyle
\int_0^\infty\,\dd x\,
x^{m}\,\exp(-a\,x)\,I_m(b\,x)\,L^{2m}_n(c\,x) \,=\,
\dfrac{b^m \Gamma(2m+1) (2m+1)_n}{2^m n! (a+b)^{2m+1} \Gamma(m+1)}\\
\\
\,=\,
\displaystyle
F_2\left(2m+1;m+\dfrac 12, -n;2m+1,2m+1;
\dfrac {2b}{a+b}, \dfrac{c}{a+b}
\right)\,,
\end{array}
\end{equation}
where the symbol $F_2$ denotes one of the so-called Appell functions~\cite[Ch. 16]{DLMF}.
In this particular case, we shall limit ourselves to real values of $(a,b,c)$ and to integer values
of $(n,m)$. Moreover, in the following it will be assumed that $a > b\ge 0$.

The Appell function $F_2$ into Eq.~(\ref{Eq:BrychkovNewFormula}) can be recast in terms of the 
Appell function $F_1$ by using the following transformation formula~\cite[formula~16.16.3]{DLMF}: 
\begin{equation}
\label{Eq:AppellF2ToAppellF1}
\begin{array}{c}
\displaystyle
F_2\left(\alpha;\beta, -n;\alpha,\alpha;x, y\right)\,=\,
(1-y)^n\,F_1\left(\beta;\alpha+n,-n;\alpha;x,\dfrac x{1-y}\right)\,,
\end{array}
\end{equation}
where $F_1(\cdot)$ denotes another member of the Appell function set.
Simple algebra then gives
\begin{equation}
\label{Eq:AppellF2ToAppellF1.1}
\begin{array}{c}
\displaystyle
F_2\left(2m+1;m+\dfrac 12, -n;2m+1,2m+1;
\dfrac {2b}{a+b}, \dfrac{c}{a+b}
\right)\,=\,\\
\\
\displaystyle
\,=\,
\left(\dfrac{a+b-c}{a+b}\right)^n\,
F_1\left(m+\dfrac 12;2m+n+1,-n;2m+1;\dfrac {2b}{a+b},\dfrac {2b}{a+b-c}\right)\,.
\end{array}
\end{equation}
Equation~(\ref{Eq:AppellF2ToAppellF1.1}) can further be simplified, by expressing the Appell
function $F_1$ in terms of the, better known Gauss hypergeometric function ${}_2F_{1}$
through the following connection formulas \cite{DLMF}:
\begin{equation}
\label{Eq:ReductionAppellF1}
\begin{array}{l}
\displaystyle
F_1\left(\beta;\alpha+n,-n;\alpha;\xi,\eta\right)\,=\,
(1-\eta)^{-\beta}\,{}_2F_{1}\left(\beta,\alpha+n;\alpha;\dfrac{\xi-\eta}{1-\eta}\right)\,=\,\\
\\
\displaystyle
\,=\,
(1-\eta)^{-\beta}\,\left(\dfrac{1-\xi}{1-\eta}\right)^{-\beta-n}\,{}_2F_{1}\left(\alpha-\beta,-n;\alpha;\dfrac{\xi-\eta}{1-\eta}\right)\,=\,\\
\\
\displaystyle
\,=\,
\left(\dfrac{1-\eta}{1-\xi}\right)^{n}\,(1-\xi)^{-\beta}\,{}_2F_{1}\left(\alpha-\beta,-n;\alpha;\dfrac{\xi-\eta}{1-\eta}\right)\,,
\end{array}
\end{equation}
which, when substituted into Eq.~(\ref{Eq:AppellF2ToAppellF1.1}) gives at once
\begin{equation}
\label{Eq:AppellF2ToAppellF1.1}
\begin{array}{l}
\displaystyle
F_2\left(2m+1;m+\dfrac 12, -n;2m+1,2m+1;
\dfrac {2b}{a+b}, \dfrac{c}{a+b}
\right)\,=\,\\
\\
\displaystyle
\,=\,
\left(\dfrac{a+b-c}{a+b}\right)^n\,
F_1\left(m+\dfrac 12;2m+n+1,-n;2m+1;\dfrac {2b}{a+b},\dfrac {2b}{a+b-c}\right)\,=\,\\
\\
\displaystyle
\,=\,
\left(\dfrac{a-b-c}{a-b}\right)^n\,
\left(\dfrac{a+b}{a-b}\right)^{m+1/2}\,
{}_2F_1\left(m+\dfrac12,-n;2m+1;
\dfrac {2bc}{(a+b)(b+c-a)}\right)\,.
\end{array}
\end{equation}
Finally, on substituting from Eq.~(\ref{Eq:AppellF2ToAppellF1.1}) into Eq.~(\ref{Eq:BrychkovNewFormula}),
after simplifying and rearranging, it is obtained
\begin{equation}
\label{Eq:BrychkovNewFormulaSimplified}
\begin{array}{l}
\displaystyle
\int_0^\infty\,\dd x\,
x^{m}\,\exp(-a\,x)\,I_m(b\,x)\,L^{2m}_n(c\,x) \,=\,\left(\dfrac b2\right)^m\,
\dfrac{(2m+n)!}{n! m!}\,
\dfrac{1}{(a^2-b^2)^{m+1/2}}\\
\\
\displaystyle
\times
\left(\dfrac{a-b-c}{a-b}\right)^n\,
{}_2F_1\left(m+\dfrac12,-n;2m+1;
\dfrac {2bc}{(a+b)(b+c-a)}\right)\,.
\end{array}
\end{equation}
Last equation can further be simplified on taking into account that
\begin{equation}
\label{Eq:BrychkovNewFormulaSimplifiedGegenbauer.1}
\begin{array}{l}
\displaystyle
{}_2F_1\left(-n,\gamma;2\gamma;z\right)\,=\,
\dfrac{n!}{\left(\gamma+\dfrac 12\right)_n}\,\dfrac 1{4^n}\,z^n\,C^{\frac 12-\gamma-n}_n\left(1-\dfrac 2z\right)\,,
\end{array}
\end{equation}
where the symbol $C^\lambda_n(\cdot)$ denotes the ultraspherical, or Gegenbauer polynomial.
Then, on substituting from Eq.~(\ref{Eq:BrychkovNewFormulaSimplifiedGegenbauer})
into Eq.~(\ref{Eq:BrychkovNewFormulaSimplified}), eventually we arrive at
\begin{equation}
\label{Eq:BrychkovNewFormulaSimplified}
\begin{array}{l}
\displaystyle
\int_0^\infty\,\dd x\,
x^{m}\,\exp(-a\,x)\,I_m(b\,x)\,L^{2m}_n(c\,x) \,=\,\\
\\
\displaystyle
\,=\,\left(\dfrac b2\right)^m\,
\dfrac{(2m+n)!}{(m+n)!}\,
\dfrac{(-bc/2)^n}{(a^2-b^2)^{n+m+1/2}}
C^{-m-n}_n\left(\dfrac{a^2-b^2-ac}{bc}\right)\,,
\end{array}
\end{equation}
which coincides with Eq.~(\ref{Eq:BrychkovNewFormulaSimplifiedGegenbauer}). 

%
%

%

\end{document}